# Electrooptic Effect in Non-Centrosymmetric CsLiB$_6$O$_{10}$ Borate Crystals

Mys O., Adamiv V., Martunyuk-Lototska I. and Vlokh R.

Institute of Physical Optics, 23 Dragomanov Str., 79005 Lviv, Ukraine

e-mail: vlokh@ifo.lviv.ua

**Abstract**

Electrooptic coefficient of CsLiB$_6$O$_{10}$ crystals has been experimentally determined as $r_{63}$=3.3×10$^{-12}$m/V. The half-wave voltage for CsLiB$_6$O$_{10}$ ($U_{\lambda/2}$=26 kV) is three orders of magnitude larger than that of the known KDP crystals.

**PACS**: 78.20.Jq, 42.60.Gd

**Key words**: electrooptic effect, half-wave voltage, borate crystals

Non-centrosymmetric borate crystals, such as β-BaB$_2$O$_4$ (point group of symmetry *3m*), Li$_2$B$_4$O$_7$ (*4mm*) and LiB$_3$O$_5$ (*mm2*), are widely used for optical harmonics generation and parametric oscillation (see, e.g., [1,2]). The advantages of application of these crystals in laser radiation controlling follow from their wide range of transparency [3], high optical damage threshold [4] and nonlinear-optic figure of merit [1]. On the other hand, these crystals could be also useful as electrooptic (EO) materials. The EO coefficients of some of the borate crystals are presented in Table 1 as an example.

Table 1. EO coefficients of borate crystals according to the literature.

| $r_{ij}$, 10$^{-12}$ m/V | β-BaB$_2$O$_4$ (group *3m*) | Li$_2$B$_4$O$_7$ (group *4mm*) |
|---|---|---|
| $r_{11}$ | 9.52 [5] | 0 |
| $r_{13}$ | - | 3.74 [6] |
| $r_{33}$ | - | 3.67 [6] |
| $r_{51}$ | 6.68 [5] | -0.11 [6] |

CsLiB$_6$O$_{10}$ crystals are quite new representatives of the borate family [7]. They belong to the point group of symmetry *42m*. It is interesting to note that the symmetry group of caesium-lithium borate is the same as that of the well-known EO material, KDP crystals. In case of EO crystals belonging to that symmetry group, the coefficient $r_{63}$ is usually utilized. The aim of the present paper is determination of that coefficient for CsLiB$_6$O$_{10}$ crystals.

The EO effect in CsLiB$_6$O$_{10}$ crystals has been studied with the Senarmont method at the laser wavelength of $\lambda$=632.8nm. The electric field has been applied along <001> axis and the light has propagated in the direction <001>.

The dependence of optical birefringence on the electric field strength presented in Figure 1 turns out to be linear.

The EO coefficient has been calculated with the formula

$$r_{63} = \frac{2\delta(\Delta n)_{12}}{n_0^3 E_3} . \qquad (1)$$

It is found to be equal to $r_{63}$=3.3 ×10$^{-12}$m/V, while the corresponding half-wave voltage is $U_{\lambda/2}$=26kV. One can see that the latter value for CsLiB$_6$O$_{10}$ crystals is three orders of magnitude greater than that characteristic of the KDP crystals [8]. Nevertheless, the damage threshold for the crystals under test is essentially higher in comparison with the KDP (26GW/cm$^2$ for CsLiB$_6$O$_{10}$ [4,9] and 17 GW/cm$^2$ for the KDP [10]). Thus, CsLiB$_6$O$_{10}$ crystals may be used for a high-power laser Q-switching.

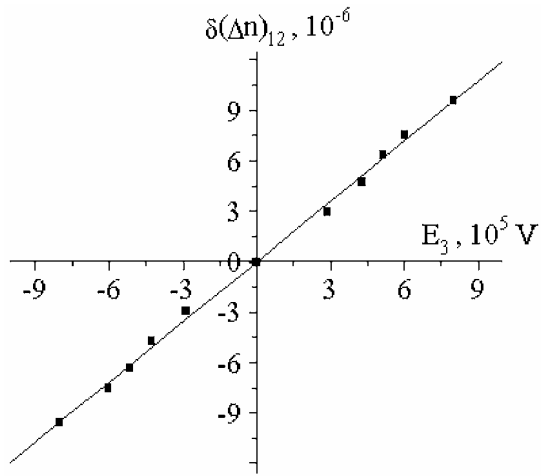

**Fig. 1.** Dependence of EO birefringence change in CsLiB$_6$O$_{10}$ crystals versus the electric field at $\lambda$=632.8 nm, T=22$^0$C.

The authors are grateful to the Ministry of Education and Science of Ukraine (the Project N0103U000703) for the financial support of this study.

**References**

1. Cheng W.-D., Huang J.-S., Lu J.-X. Phys. Rev. B. **57** (1998) 1527.
2. Whatmore R.W., Shorrocks N.H., O'Hara C., Alinger F.W. Electron. Lett. **17** (1981) 11.
3. Martynyuk-Lototska I., Mys O., Krupych O, Adamiv V., Burak Ya., Vlokh R., Schranz W. Ferroelectrics. (at press).
4. Vlokh R., Dyachok Ya., Krupych O., Burak Ya., Martunyuk-Lototska I., Andru-shchak A., Adamiv V. Ukr. J. Phys. Opt. **4** (2003) 101.
5. Bohaty L., Haussuhl S., Liebertz J. Cryst. Res. Technol. **24** (1989) 1159.
6. Adamiv V.T., Burak Ya.V., Dovgii Ya.V., Kityk I.V. Zh. Prikl. Spektrosk. **54** (1994) 92 (in Russian).
7. Sasaki T., Mori Y., Kuroda I., Nakayama S., Yamaguchi K., Watanabe S. Acta Cryst. C **51** (1995) 2222.
8. Zwiker B., Scherrer P. Helv. Phys. Acta **16** (1943) 214; Zwiker B., Scherrer P. Helv. Phys. Acta **17** (1944) 346.
9. Kokh A.E, Kononova N.G., Lisova I.A., Muraviov S.V. Proc. SPIE **4268** (2001) 43.
10. Nakatani H., Bosenberg L.K., Cheng, Tang T.A. Appl. Phys. Lett. **53** (1988) 2587.